\documentclass[12pt]{article}
\usepackage{graphicx}
\begin{document}
\title{Nonhomogeneity driven Universe acceleration}
\author{Jerzy Stelmach ~and Izabela Jakacka\\
Institute of Physics, University of Szczecin,\\
 Wielkopolska 15, 70-451 Szczecin, Poland}
\date{\today}
\maketitle

E-mail: {\tt jerzyst@wmf.univ.szczecin.pl}

and~~~~~ {\tt ijakacka@poczta.onet.pl}

\begin{abstract}
Class of spherically symmetric Stephani cosmological models is examined
in the context of evolution type. It is assumed that the equation of
state at the symmetry center of the models is barotropic $(p(t)=
\alpha\rho(t))$ and the function $k(t)$ playing role of spatial curvature
is proportional to Stephani version of the Friedmann-Robertson-Lemaitre-Walker
scale factor $R(t)$ $(k(t)=\beta R(t)).$

Classification of cosmological models is performed depending on different
values and signs of parameters $\alpha$ and $\beta$. It is shown that for
$\beta < 0$ (hyperbolic geometry) dust-like $(\alpha=0)$ cosmological
model exhibits accelerated expansion at later stages of evolution.

The Hubble and deceleration parameters are defined in the model and it is shown
that the deceleration parameter decreases with the distance becoming negative
for sufficiently distant galaxies.

Redshift-magnitude relation $m(z)$ is calculated and discussed in the context of
SnIa observational data. It is noticed that the most distant supernovae of
type Ia fit quite well to the relation $m(z)$ calculated in the considered model
($H_0=65$ km/sMpc, $\Omega_0\leq 0.3$) without introducing the cosmological constant.

It is also shown that the age of the universe in the model is longer than in the
Friedmann model corresponding to the same $H_0$ and $\Omega_0$ parameters.
\end{abstract}
\vspace{2mm}

\section{Introduction}

In recent years strong observational evidencies appeared supporting the
claim that the universe is accelerating its expansion. The observations
are based on supernovae of type Ia (SnIa) used as standard candles in
order to calculate the distance to them \cite{riess98}. Since in standard
Friedmann-Robertson-Lemaitre-Walker (FRLW) cosmological models, with
nonrelativistic matter and radiation, acceleration is not possible, some
people argue that the universe must be dominated by cosmological constant
or by some other smooth exotic component influencing negative pressure.
There are many different physical scenarios trying to explain existence
of such exotic form of matter \cite{vil84,dav87,silv94,kam96,cald98}.

In the present paper we propose different approach, in which exotic matter
is not necessary to drive accelerated expansion. Instead we relax the
assumption of homogeneity of space leaving the isotropy with respect to
one point. In other words we replace the usual Cosmological Principle
by Copernican Cosmological Principle \cite{rud95} or some
generalized version of the Copernican Principle stating that our
Galaxy is at the centre of the Universe. From philosophical point
of view this assumption seems to be very artificial but our
intention is to show that even such an exotic model does not
contradict observations \cite{ellis78}.

We shall deal with spherically symmetric
Stephani models \cite{stepha67,stephb67,kramer80,kras83,kras97,dab93}
with the metric given by
\begin{equation}
ds^2=c^2\left[F(\tau) {R(\tau)\over V(r,\tau)} {d\over d\tau}
\left(V(r,\tau)\over R(\tau)\right)\right]^2d\tau^2- {R^2(\tau)\over
V^2(r,\tau)} (dr^2+r^2d\Omega^2),
\end{equation}
where the functions $V(r,\tau)$ and $F(\tau)$ are defined
\begin{eqnarray}
V(r,\tau)&=& 1+ {1\over 4} k(\tau)r^2,\\
F(\tau)&=& {1 \over c} {R(\tau) \over
\sqrt{C^2(\tau)R^2(\tau)-k(\tau)}}.
\end{eqnarray}
The functions $C(\tau), k(\tau)$ and $R(\tau)$ are not all independent but
are related with each other with the help of expressions
\begin{eqnarray}
\rho(\tau)&=& {c^4\over 8\pi G} 3C^2(\tau),\\
p(r,\tau)&=&{c^4\over 8\pi G}\left[2C(\tau)\dot C(\tau) {V(r,\tau)/R(\tau)
\over (V(r,\tau)/R(\tau))\dot{}} - 3C^2(\tau)\right],
\end{eqnarray}
following from the Einstein equations. $(~~)\dot{}$ denotes a derivative
with respect to $\tau$. Note that in the considered spherically symmetric
Stephani models and in the given coordinate system, the energy density
$\rho(\tau)$ is uniform, while the
pressure $p(r,\tau)$ is not and depends on the distance from the symmetry
center placed at $r=0$. This is the reason why in such models barotropic
equation of state (i.e. of the form $p=p(\rho))$ does not exist. If we,
however, assume some relation between $\rho(\tau)$ and $p(r,\tau)$, this
could allow us to eliminate one of the unknown functions, e.g. $C(\tau)$.
Hence we are left with two unknown functions $k(\tau)$ and $R(\tau)$. The
first one $k(\tau)$ plays a role of spatial curvature index, while the
second one $R(\tau)$ is Stephani version of FRLW scale factor. As far as
in FRLW models the curvature index is constant $(k=0,\pm 1)$ and does not
change in the process of evolution of the universe, in Stephani models
this is not the case. Due to matter transfer, caused by the gradient of
pressure, the curvature index may change with time \cite{bar00}. Similarly
as in FRLW cosmology, where the value of $k$ must be determined from
observations, in Stephani models the choice concerns the whole function
$k(\tau)$. As regards the scale factor $R(\tau)$, some arbitrariness is
connected with the choice of  time parameter $\tau$. We remove this
arbitrariness by assuming that the time parameter is a proper time of an
observer placed at the symmetry center of the model $(r\approx 0)$.

In the next section we determine spherically symmetric Stephani cosmological
model by assuming barotropic equation of state in the neighbourhood of
the symmetry center and by assuming a relation between the scale factor
$R(\tau)$ and arbitrary function $k(\tau)$.

In Section 3 we perform classification of possible cosmological models
depending on two parameters: $\alpha$-parameter appearing in the equation
of state, therefore determining kind of matter filling up the universe,
and $\beta$-parameter responsible for the global geometry of the model.

In Section 4 we focus our attention on one model corresponding to vanishing
$\alpha$-parameter (dust-like matter) and to negative $\beta$-parameter
(hyperbolic geometry). The model exhibits accelerated expansion at later
stages of evolution.

Section 5 is devoted to definition of Hubble and deceleration parameters in
spherically symmetric Stephani models. We show that in the considered model,
in spite of the fact that local observations of galaxies give positive
deceleration parameter, in the case of more distant ones, negative
deceleration parameter can be observed.

In Section 6 we discuss the redshift-magnitude relation in the model and fit
our theoretical curves to the supernovae Ia observational data.

In Section 7 we compare age of the universe in the considered Stephani
model to the age in FLRW model determined by the same $H_0,~\Omega_0$
and $\alpha$ parameters.

In Section 8 we summarize the results.

\section{Determination of the model}

Now we consider an observer placed at the symmetry center of the
spherically symmetric Stephani Universe $(r\approx 0)$. All our physical
assumptions will concern his neighbourhood.

First of all we assume that locally, matter filling up the Universe
fulfills a barotropic equation of state of the standard form
\begin{equation}
p(r\approx 0,\tau)=\alpha\rho(\tau),
\end{equation}
where $\alpha$ is some constant $(\alpha\geq -1)$. For example $\alpha=0$
corresponds to dust and $\alpha=1/3$ to relativistic matter. For negative
$\alpha$ pressure becomes negative and we regard an appropriate matter to
be in exotic form. An extreme value of $\alpha$ $(\alpha=-1)$ corresponds
to cosmological constant.

Substituting (4) and (5) into (6) we find an explicit form of the function
$C(\tau)$
\begin{equation}
C(\tau)={A\over R^{3(1+\alpha)/2}(\tau)},
\end{equation}
where $A$ is some constant. Choosing a curvature function $k(\tau)$ in a
form
\begin{equation}
k(\tau)=\beta R(\tau)
\end{equation}
determines uniquely the function $F(\tau)$, which now reads
\begin{equation}
F(\tau)= {1\over c}{R^{3(1+\alpha)/2}(\tau) \over \sqrt{A^2-\beta R^{2+3\alpha}
(\tau)}}.
\end{equation}
$\beta$ is a constant, which can be either negative, zero or positive.

The condition (8) is of course not the most general one. The 
 simplicity was the only criterion for choosing it. However,
one can easily show that any condition of the form $k(\tau)=\beta R^\gamma(\tau)$
($0<\gamma<2$) gives similar results to that ones obtained in the
present paper.

Now we are left with the time parameter $\tau$ which is not uniquely
determined so far. In FRLW models the time parameter is usually chosen
globally to be a proper time of any comoving observer. In
spherically symmetric Stephani models usually different choice is realized.
Since there exists only one symmetry center in the models it seems natural
to distinguish just this point. Hence the time parameter is chosen in a
special way to be a proper time of an observer placed at the symmetry
center. It is not a proper time of any other observer. To do
that we put \cite{dab95}
\begin{equation}
F(\tau)R(\tau){d\over d\tau} \left[V(r,\tau)\over R(\tau)\right]=1,
\end{equation}
what together with (8) and (9) gives a dynamical equation for the
Stephani version of the scale factor $R(t)$
\begin{equation}
\left(dR(t) \over cdt\right)^2+\beta R(t)={A^2 \over R^{1+3\alpha}(t)},
\end{equation}
or more familiar form resembling a Friedmann equation
\begin{equation}
\left(dR(t)\over cdt\right)^2+k(t)={8\pi G \over 3c^4} \rho(t)R^2(t),
\end{equation}
where the energy density $\rho(t)$ is calculated from (4)
\begin{equation}
\rho(t)={C_\alpha \over R^{3+3\alpha}(t)},
\end{equation}
and $C_\alpha$ is some constant. New time parameter is now denoted by $t$
and will be regarded as a cosmic time. Solution of the equation (11)
could give us a full information about evolution of the model. What we
are interested in is how the type of evolution is influenced by the
numerical values of parameters of $\alpha$ and $\beta$. Note that
Barrett and Clarkson \cite{bar00} also discussed two parameter
class of spherically symmetric Stephani models but of different
type. Their models do not admit dust-like $(\alpha=0)$ equation of
state at the symmetry center.

\section{Classification of models}

Qualitative classification of spherically symmetric cosmological models
depending on different values of parameters $\alpha$ and $\beta$ can be
performed without solving the equation (11). For this purpose we use a
standard method \cite{coq82,dab86} consisting in treating the evolution
equation (11) as the energy conservation principle for some dynamical
system with kinetic energy $(dR/cdt)^2$ and potential energy
\begin{equation}
V(R)=\beta R-{A^2\over R^{1+3\alpha}}.
\end{equation}
Then the energy conservation principle reads
\begin{equation}
\left(dR \over cdt \right)^2+V(R)=0.
\end{equation}
Now, illustrating the potential energy $V(R)$ on a diagram and keeping in
mind that $(dR/cdt)^2$ must be positive we deduce qualitative behaviour of
the scale factor $R(t)$ for different values of $\alpha$ and $\beta$. We
do not consider the case $\beta=0$ since it corresponds to FRLW
Einstein-de Sitter model. All other models split naturally into two cases
depending on the sign of the $\beta$-parameter. Moreover in the frame of
each case five subcases can be pointed out, corresponding to the
following ranges of the $\alpha$-parameter:

\par a) $\alpha > -1/3$,
\par b) $\alpha = -1/3$,
\par c) $-2/3<\alpha < -1/3$,
\par d) $\alpha = -2/3$,
\par e) $-1\leq\alpha < -2/3$.

\begin{figure}
\begin{center}
\includegraphics[angle=0, width=0.65\textwidth]{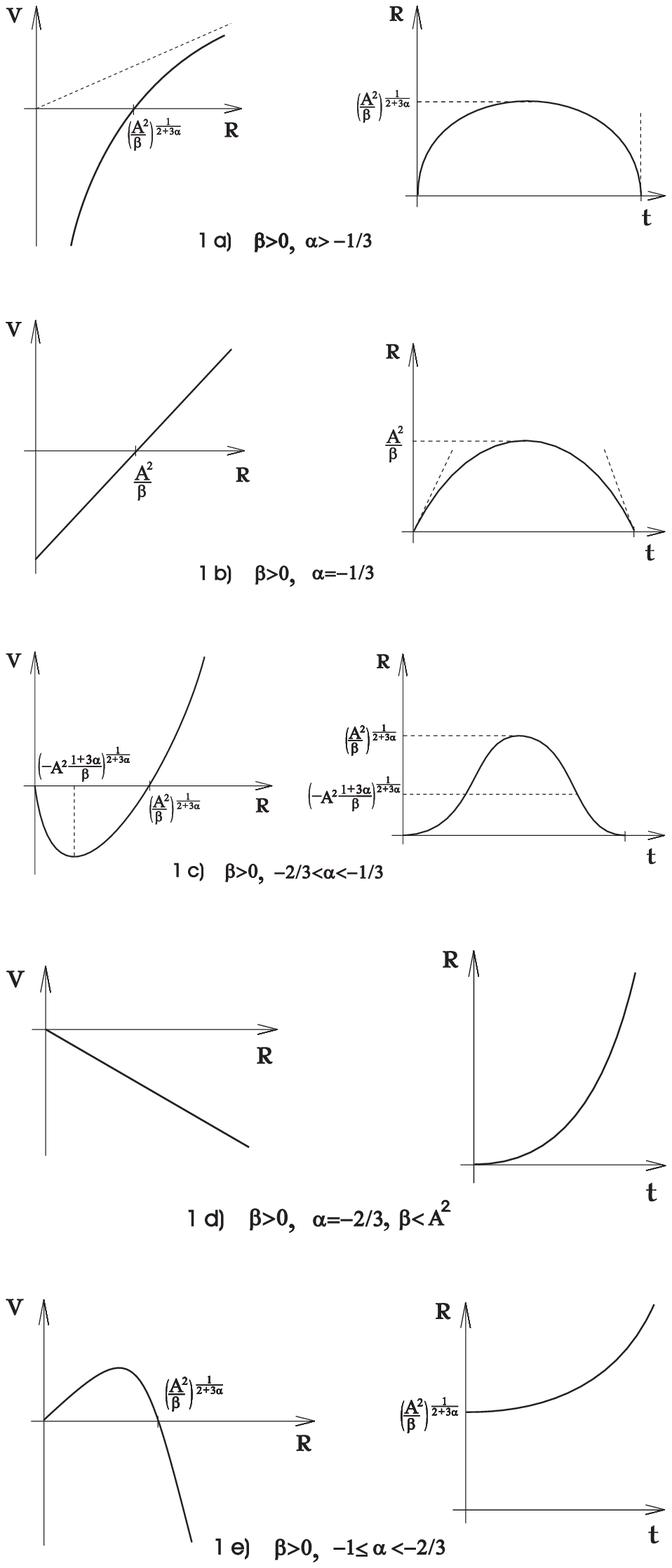}
\caption{Potential energy $V(R)$ of associated dynamical system (left
diagrams) and corresponding evolution of the scale factor $R(t)$ (right
diagrams) for positive $\beta$-parameter (closed geometry) and different
values of $\alpha$-parameter.}
\label{fig:1}
\end{center}
\end{figure}

\begin{figure}
\begin{center}
\includegraphics[angle=0, width=0.62\textwidth]{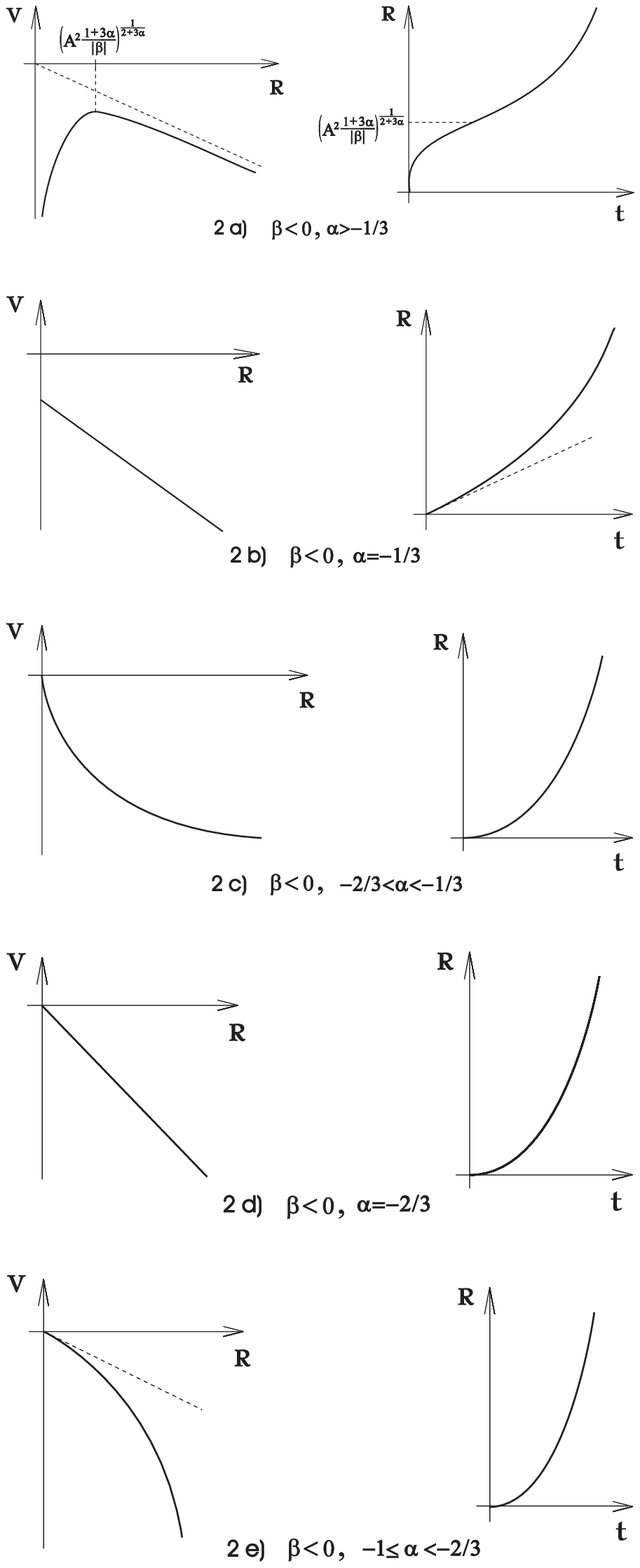}
\caption{Potential energy $V(R)$ of associated dynamical system (left
diagrams) and corresponding evolution of the scale factor $R(t)$ (right
diagrams) for negative $\beta$-parameter (open geometry) and different
values of $\alpha$-parameter.}
\label{fig:2}
\end{center}
\end{figure}

Appropriate behaviours of the potential energy $V(R)$ and corresponding
evolution of the scale factor $R(t)$ are illustrated in Figs 1 and 2.

Some of the above presented cosmological models were already discussed in
detail in literature. Especially much attention was paid to the model
corresponding to $\alpha=-1/3$ (Figs 1b, 2b) due to its particular
simplicity (scale factor can be calculated explicitly) \cite{dab95,dab99}.
In fact, some other models can also be expressed in an analytic way. In the
present paper, however, we shall concentrate on the case, which
mathematically is not so simple, but from physical point of view is very
interesting because it relates to the SnIa observations suggesting
accelerated expansion of the Universe at the present epoch \cite{riess98}.
The model we want to discuss is presented in Fig. 2a.

\section{Accelerating Universe}

It is well known that in FRLW cosmology accelerated expansion of the
universe may be only obtained in the case when the universe is filled with
some exotic form of matter influencing negative pressure, e.g. cosmological
constant. However, if we look at Fig. 2a we notice that in the spherically
symmetric Stephani model the universe can exhibit accelerated expansion
also in the case of vanishing or positive pressure at least at the
symmetry center. Strictly speaking accelerated expansion takes place for
every value of $\alpha$ at least at later stages of evolution, provided
that $\beta<0$ (hyperbolic geometry). But it seems to us that the model
corresponding to vanishing $\alpha$ is the simplest one, moreover it
describes locally (in the neighbourhood of the observer located at the
symmetry center) pressureless matter-dominated universe which is in
agreement with local observations.

So putting $\alpha=0$ we find the energy density $\rho(t)$ and the
pressure $p(r,t)$ to be equal to
\begin{equation}
\rho(t) = {3c^4A^2 \over 8\pi G} {1\over R^3(t)}
\end{equation}
and
\begin{equation}
p(r,t)= {3c^4A^2 \over 32\pi G}\beta r^2 R(t).
\end{equation}
Note that indeed at the symmetry center $(r\approx 0)$ pressure vanishes,
however, due to hyperbolic geometry $\beta<0$ it effectively becomes
negative at larger distances from the symmetry center. So in spite of the
fact that the energy density scales with the expansion as ordinary
nonrelativistic matter $\rho\propto R^{-3})$, due to negativeness of
pressure, for $r>0$, it is exotic. Stephani version of the Friedmann
equation now reads
\begin{equation}
\left(dR\over cdt\right)^2+\beta R={A^2\over R}
\end{equation}
and unfortunately cannot be solved in an elementary way. However in two
extreme cases corresponding to small and large $R$ exact solutions can
be easily found. They are
\begin{equation}
\nonumber
R(t)=\left({3\over 2}Ac\right)^{2/3}t^{2/3}~~~~{\rm for~ small}~~ R,
\end{equation}
and
\begin{equation}
R(t)={1\over 4}c^2\vert\beta\vert t^2~~~~{\rm for~ large}~~R.
\end{equation}
In the first case the solution is of Einstein-de Sitter type, while in
the second case the evolution is evidently inflationary of power law
type. It means that in our model the universe starts as
decelerating and finally ends up as accelerating one. In the
simplest FRLW cosmological models with one component fluid filling
up the universe such behaviour is not possible. In the present
model, inspite of the fact that formally it is of one component
type (dust-like matter) the curvature term is non-trivial (see the
condition (8)) and simulates the existence of some exotic fluid
driving the power-law inflation at later epoch.

In the next Section we discuss some observational parameters in the model.

\section{Observational Parameters}

Observational parameters which describe kinematics of the evolution of
the universe are Hubble parameter $H$ and deceleration parameter $q$. In
FLRW models they are constant on time slices due to homogeneity and
isotropy. It does not to be the case in spherically symmetric Stephani
models, in which homogeneity does not take place.

Following Ellis \cite{ellis73} we define $H$ and $q$ in a general case by
\begin{equation}
H\equiv {1\over l}{dl\over d\tau}, ~~~~q\equiv -\left({dl\over d\tau}
\right)^{-2}\left({d^2l\over {d\tau}^2}\right)l,
\end{equation}
where $\tau$ is here proper time measured along the particle world line,
and $l$ is some representative length along the particle world-lines.

In FLRW cosmological models the above expressions go over into the
standard ones
\begin{equation}
H={\dot R\over R}, ~~~~q=-{\ddot R R\over {\dot R}^2},
\end{equation}
where $R$ is a scale factor entering the space part of the line-element
\begin{equation}
ds^2=c^2d\tau^2-R^2(\tau)\left({dr^2\over 1-kr^2}+r^2d\Omega^2\right).
\end{equation}
In our spherically symmetric Stephani models the line element is given by
\begin{equation}
ds^2={c^2dt^2\over V^2(r,t)}-{R^2(t)\over V^2(r,t)}\left(dr^2+r^2d
\Omega^2\right).
\end{equation}
Hence neither $t$ is proper time of any comoving observer, nor $R(t)$ is
a true scale factor (uniquely determining the length $l$). In order to
calculate $H$ and $q$ we have to rewrite the line element using global
time parametrization
\begin{equation}
ds^2=c^2d\tau^2-R^2(r,\tau)\left(dr^2+r^2d\Omega^2\right),
\end{equation}
where we defined
\begin{equation}
d\tau\equiv{dt\over V}, ~~~~R(r,\tau)\equiv{R(t(r,\tau))\over
V(r,t(r,\tau))}.
\end{equation}
Now the definitions for $H$ and $q$ read
\begin{equation}
H={1\over R(r,\tau)}{dR(r,\tau)\over d\tau}, ~~~~q=-{1\over H^2R(r,\tau)}
{d^2R(r,\tau)\over d\tau^2}.
\end{equation}
Note that generally $H$ and $q$ are not spatially constant anymore, what
should not be surprising in non-homogeneous universe. It can be shown,
however, that it is possible to choose time parameter with the aid of
which defined Hubble parameter does not depend on $r$. It turns out that
the time parameter $t$ in the line element (24) is just such a
parameter, i.e.
\begin{equation}
H={1\over R(r,\tau)}{dR(r,\tau)\over d\tau}={1\over R(t)}{dR(t)\over dt}.
\end{equation}
In other words with this time parameter the expansion is the same at
every point of time slice of the model. Unfortunately it does not happen
in the case of $q$ parameter which is evidently coordinate dependent
\cite{suss99}. Explicit calculation gives
\begin{equation}
q(r,t)=-V(r,t){\ddot R(t)\over R(t)H^2}+V(r,t)-1=-V(r,t)q(t)+V(r,t)-1
\end{equation}
or
\begin{equation}
q(r,t)=q(t)+{1\over 4}\beta r^2R(t)(1+q(t)),
\end{equation}
where we defined
\begin{equation}
q(t)=-{\ddot R(t)\over R(t)H^2(t)}.
\end{equation}
for deceleration parameter in neighbourhood of the symmetry center.

Since $\beta<0$ in our model, the deceleration parameter $q(r,t)$
decreases with increasing distance to the observed galaxy. Physically it
means that in our non-homogeneous universe acceleration of the expansion
increases with distance.

In order to see how the deceleration parameter $q(r,t)$ depends on the
redshift we express $\beta,~r$ and $R(t)$ in terms of observational
parameters $H_0,~\Omega_0$ and $z$, where $\Omega_0$ is energy density
parameter at present epoch and $z$ is a redshift of a galaxy placed at
some point determined by comoving coordinate $r$. Since in the considered
Stephani models energy density $\rho(t)$ is constant in space, energy
density parameter $\Omega$ can be defined similarly as in FLRW models, i.e.
\begin{equation}
\Omega(t)={\rho(t)\over \rho_{cr}(t)},
\end{equation}
where
\begin{equation}
\rho_{cr}(t)\equiv {3c^2\over 8\pi G}H^2(t)\label{eq:b}
\end{equation}
is critical energy density defined as actual energy density in the model
with flat geometry $(\beta=0).$

Expressing the comoving coordinate $r$ in terms of observational
parameters is essential for our purposes and can be performed in a
standard way by putting the infinitesimal space-time interval $ds^2$
connecting two light events equal to zero (we choose coordinate system in
such a way that $\theta$ and $\phi$ are constant along light ray path from
galaxy to observer). Hence we get
\begin{equation}
cdt=-R(t)dr
\end{equation}
or equivalently
\begin{equation}
{R^{(3\alpha-1)/2}dR\over \sqrt{A^2-\beta R^{2+3\alpha}}}=-dr
\end{equation}
after getting rid of $t$ variable using (11). Constants $\beta$ and $A^2$
can be found by writing down Friedmann equation (11) or (12) for the present
epoch $(t=t_0)$ and by making use of the definition of $\Omega_0$
parameter:
\begin{eqnarray}
\beta&=&{1\over c^2}R_0{H_0}^2(\Omega_0-1),\\
A^2&=&{1\over c^2}{R_0}^{3+3\alpha}{H_0}^2\Omega_0.
\end{eqnarray}
Substituting these expressions into (35), next integrating both sides and
changing integration variable we get integral formula for $r$
\begin{equation}
r={c\over R_0 H_0}\int_{R/R_0}^1\limits{x^{(3\alpha-1)/2}dx\over
\sqrt{\Omega_0+(1-\Omega_0)x^{2+3\alpha}}}.
\end{equation}
$R$ is a scale factor at the epoch when the light ray was emitted by the
observed galaxy. Unfortunately integration cannot be performed explicitly
either in general case or even in the special case corresponding to
$\alpha=0.$ We compute it approximately assuming that $\Omega_0$ is small
with respect to the ratio $R/R_0.$ Physically it means that in order to
keep the approximation valid, the larger distances $r$ are calculated,
the smaller energy density parameter $\Omega_0$ has to be taken into account.
Then we skip the first term under the square root in (38) and the integration
can be carried out explicitly giving
\begin{equation}
r={2c\over R_0 H_0\sqrt{1-\Omega_0}}\left(\sqrt{R_0\over R}-1\right).
\end{equation}
Together with the relation \cite{dab95}
\begin{equation}
{R_0\over R}={z+1\over V(r,t)}
\end{equation}
we obtain system of algebraic equations which can be solved with respect
to $R$ and $r$
\begin{eqnarray}
R&=&{4R_0\over (z+2)^2},\\
r&=&{cz\over R_0 H_0\sqrt{1-\Omega_0}}.
\end{eqnarray}
Substituting (28a) and (33) into (24b) we find present value $(t=t_0)$ of
the deceleration parameter of the region of the universe where a galaxy
with the observed redshift $z$ is placed
\begin{equation}
q(z,t_0)=q_0-{1\over 4}(1+q_0)z^2.
\end{equation}
$q_0$ is a local value of the deceleration parameter.

In this model, in spite of the fact that local observations of galaxies
give positive value of $q_0$, in the case of more distant ones
$(z>\sqrt{q_0/(1+q_0)})$ negative deceleration parameter is measured.

More detailed description of the model from the point of view of
observations requires derivation of coordinate independent relation
between observational parameters. Redshift-magnitude relation is just
such a relation and finding it will be our next task.

\section{Redshift-magnitude relation}

In cosmological models which are not homogeneous and isotropic one of the
methods allowing for finding redshift-magnitude relation is based on
formalism proposed by Kristian and Sachs \cite{kris66} and Ellis and
MacCallum \cite{ellis70} leading to relation in form of power series
around the observer's position. But this method is not very effective if
large redshifts are taken into account.

Due to spherical symmetry of the model it is possible, however, to find
closed formula for apparent bolometric magnitude $m$ of galaxy, as a
function of redshift $z$ in the case of small $\Omega_0$-parameter. The
problem consists then in explicit calculation of the luminosity distance $D_0$.

The formula for the luminosity distance in our spherically symmetric model is 
the same as in FLRW models and reads \cite{part84}
\begin{equation}
D_0=R_0 r(z+1).
\end{equation}
Making use of earlier derived expressions for $\beta,~r$ and $R$ we find
\begin{equation}
D_0={c\over H_0}z{(z+2)^2\over 4}.
\end{equation}
For small redshifts $(z\ll 1)$ the expression goes over into usual
distance-redshift relation known from Friedmann cosmology (we remind that
the above formula is valid only for small values of $\Omega_0$)
\begin{equation}
D_0\approx{c\over H_0}z.
\end{equation}
However, for larger redshifts it grows faster than in FLRW models, where
\begin{equation}
D_0={2c\over H_0{\Omega_0}^2}[\Omega_0z+(\Omega_0-2)(\sqrt{\Omega_0z+1}-1)],
\end{equation}
and is in favor of SnIa observations.

Finding general formula for $D_0$ requires numerical integration of (38).
Before doing that we expand $D_0(z)$ in our model into Taylor series
with respect to the $z$-variable. Up to the second order we get (for arbitrary
$\Omega_0$ and $\alpha$)
\begin{equation}
D_0(z)={c\over H_0}\left\{z+{z^2\over 4}\left[4-3\Omega_0(\alpha+1)\right]\right\}.
\end{equation}
Comparing with the analogous expansion of the Mattig formula \cite{mattig58},
which up to the same order in the dust filled universe ($\alpha=0$) reads
$D_0(z)=(c/H_0)[z+(2-\Omega_0)z^2/4]$, we notice that for arbitrary
$\Omega_0$ redshift-magnitude formula in the Stephani model indeed
grows faster than in the Friedmann model at least for relatively small values of $z$,
for which the Taylor expansion is valid.

Since supernovae of type Ia favoring accelerated evolution of the Universe were
discovered at relatively large redshifts $z>0.5$ full formula for
the luminosity distance is required valid not only for small $z$.
Hence Taylor expansion calculated up to low order is not satisfactory and we must
calculate $D_0(z)$ numerically.

The results of this numerical calculation are presented in figures 3 and 4. First 
(figure 3), we compare the redshift magnitude relation obtained in both Friedmann and Stephani
spherically symmetric models determined by the same cosmological parameters
$H_0, \Omega_0$ and the $\alpha$-parameter responsible for the form of matter
filling locally the universe. We realize that for relatively large redshifts the
curves $m(z)$ in both models diverge and the relation $m(z)$ in the
Stephani model looks like the appropriate relation in the Friedmann model but
with the cosmological constant. In other words nonhomogeneity in the spherical
symmetric Stephani model in some sense simulates existence of the $\Lambda$-term.

\begin{figure}
\begin{center}
\includegraphics[angle=0, width=0.69\textwidth]{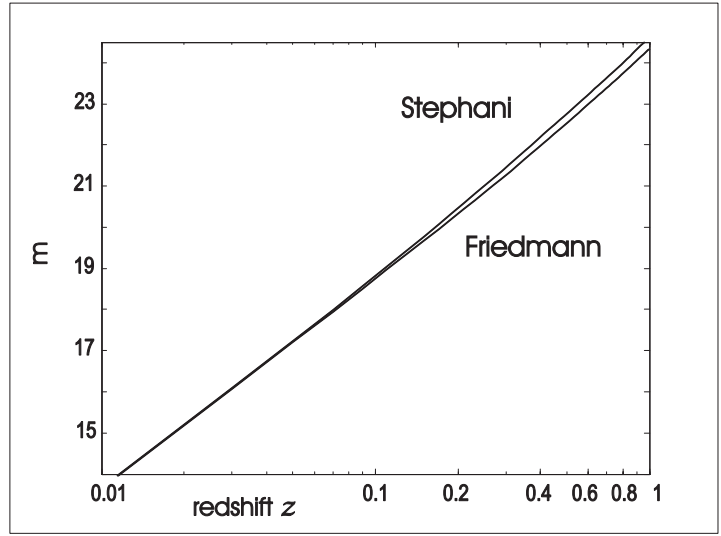}
\caption{Redshift-magnitude relation $m(z)$ in Friedmann cosmological model
(lower line) and in the spherically-symmetric Stephani model (upper line).
In both models the same cosmological parameters are chosen: $H_0=65$ km/sMpc,
$M=-19.5, \Omega_0=0.3$.}
\label{fig:3}
\end{center}
\end{figure}

In Fig. 4 the redshift-magnitude relation in the Stephani model is compared with
some observational data from the Supernova Cosmology Project \cite{perl98} (see
also \cite{hamuy96}). We notice that the most distant supernovae of type Ia fit
quite well to the dust-like ($\alpha=0$) Stephani cosmological model with the
energy density parameter $\Omega_0\leq 0.3$. The values of the Hubble constant
$H_0$ and the absolute magnitude $M$ are the same as in \cite{perl98, dab98}.

\begin{figure}
\begin{center}
\includegraphics[angle=0, width=0.69\textwidth]{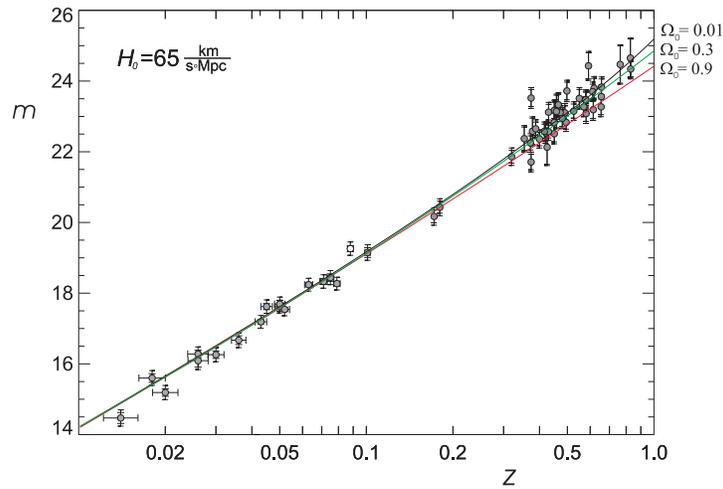}
\caption{Redshift-magnitude relation $m(z)$ fitted to observational data
from the Supernova Cosmology Project. $H_0=65$ km/sMpc,
$M=-19.5, \Omega_0=0.3$.}
\label{fig:4}
\end{center}
\end{figure}

In the next section we compare numerically the age of the universe in the
considered model and in FLRW cosmological models corresponding to the
same value of $H_0$ and $\Omega_0$ parameters.

\section{Age of the universe}

We find the age of the universe in the considered models by direct
integration of Stephani version of the Friedmann equation (11)
\begin{equation}
{t_0}^S={1\over c}\int_0^{R_0}\limits{R^{(1+3\alpha)/2}dR\over \sqrt{A^2-\beta
R^{2+3\alpha}}}={1\over H_0}\int_0^1\limits{x^{(1+3\alpha)/2}dx\over
\sqrt{\Omega_0+(1-\Omega_0)x^{2+3\alpha}}}.
\end{equation}
Analogously calculated age of the universe in FLRW models reads
\begin{equation}
{t_0}^F={1\over H_0}\int_0^1\limits{x^{(1+3\alpha)/2}dx\over
\sqrt{\Omega_0+(1-\Omega_0)x^{1+3\alpha}}}.
\end{equation}
We easily realize that the age in Stephani models is longer than in FLRW
ones corresponding to the same values of the parameters $H_0,~\Omega_0$
and $\alpha$. For $\alpha=0$ and $H_0=100h$ km/sMpc ($1/2\leq h\leq 1$)
and on the same diagram (Fig. 5) we plot the age of the universe,
${t_0}^S$ and ${t_0}^F$, in both models as a function of energy density
parameter $\Omega_0$.

\begin{figure} 
\begin{center}
\includegraphics[angle=0, width=0.55\textwidth]{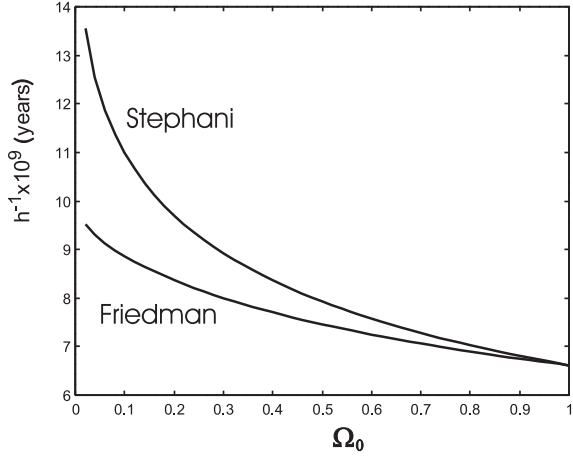}
\caption{Age of the universe in  spherically symmetric Stephani
cosmological model (upper line) and in FLRW cosmological model (lower
line) as a function of energy density parameter $\Omega_0$. It is assumed
that $\alpha=0$ and $H_0=100h$ km/sMpc, where $1/2\leq h\leq 1$.}
\label{fig:5}
\end{center}
\end{figure}

In Fig. 6 we plot the age ${t_0}^S(\alpha, \Omega_0)$ of the Universe in
Stephani model as a function of two parameters: $\alpha$ -- determining
equation of state at the symmetry center, and $\Omega_0$ -- energy density
parameter. We notice that the age increases with decreasing both $\alpha$
and $\Omega_0$ (provided that $\alpha\geq -2/3$). In Fig. 7 we present
sections of the plot ${t_0}^S(\alpha, \Omega_0)$ corresponding to the
values of $\alpha=-2/3,~-1/3,~0,~1/3,~2/3,~1$.

\begin{figure}
\begin{center}
\includegraphics[angle=0, width=0.71
\textwidth]{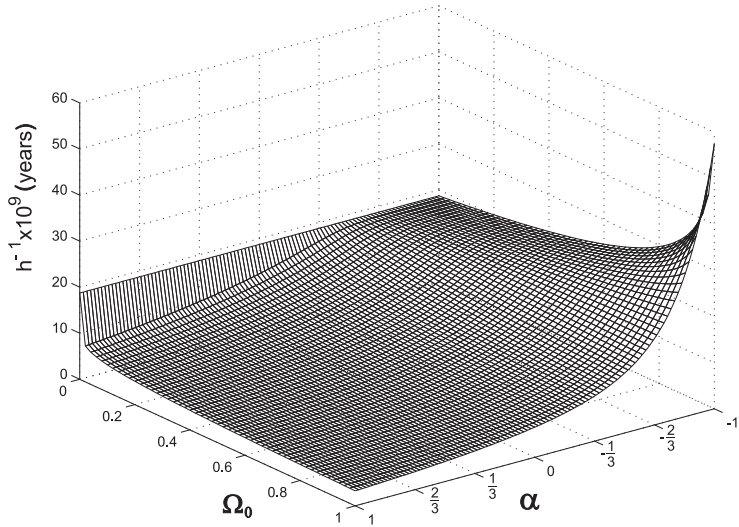}
\caption{Age of the universe in spherically symmetric Stephani
cosmological model as a function of $\alpha$ and $\Omega_0$.}
\label{fig:6}
\end{center}
\end{figure}

\begin{figure}
\begin{center}
\includegraphics[angle=0, width=0.71\textwidth]{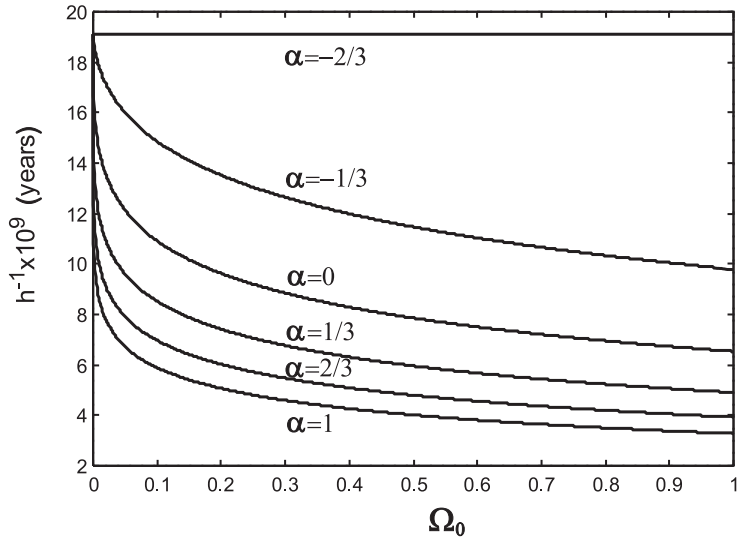}
\caption{Age of the universe in spherically symmetric Stephani
cosmological model as a function of $\Omega_0$ for different values
of $\alpha=-2/3,~-1/3,~0,~1/3,~2/3,~1$.}
\label{fig:7}
\end{center}
\end{figure}

\section{Conclusions}

In the paper we considered a class of spherically symmetric Stephani
cosmological models parametrized by two parameters $\alpha$ and $\beta$.
The $\alpha$-parameter determines the form of barotropic equation of
state in the neighbourhood of the symmetry center of the models
and $\beta$-parameter is related to the curvature index. We performed
qualitative classification of evolution types of models depending on
different values of $\alpha$ and $\beta$. We showed that for negative
$\beta$ the universe exhibits accelerated expansion independently of
$\alpha$, what does not take place in FLRW models. We focused our
attention on the model corresponding to $\alpha=0$. In this model
pressure is negligible in the neighbourhood of the symmetry center (the
so-called dust-like model) and the model looks like of Einstein-de
Sitter type at the early stage of evolution. At later epoch the
evolution is inflationary of power law form. This result seems to be
interesting because we showed that existence of ``exotic'' matter
(influencing negative pressure) is not necessary (at least locally) to
drive accelerated expansion of more distant regions. We also showed that
acceleration becomes larger while increasing the distance to the galaxy.

We derived redshift-magnitude formula in the model and showed that for
large relatively large redshifts ($z\approx 1$) apparent bolometric magnitude 
$m(z)$ grows faster than in
corresponding FLRW model what is in favor of SnIa observational
data. We found numerically that a good fit is obtained for $H_0=65$ km/sMpc
and $\Omega_0\leq 0.3$. At this point it should be noted that the good fit of
$m(z)$ to the SNIa observational data in spherically-symmetric Stephani
cosmological model has been already obtained by D\c abrowski and Hendry
\cite{dab98}. However in their model they assumed a string-like ($\alpha=-1/3$)
equation of state at the center of symmetry.

Finally we found that the age of the universe in the considered model is
remarkably larger than in isotropic and homogeneous models.

Concluding we would like to stress that we do not claim that our surrounding
universe is of Stephani type. We wanted to show that there exist
different from isotropic and homogeneous cosmological models which give
the same observational evidencies as FRWL models with the cosmological constant.
In these models the deviation from homogeneity could be so small that locally
(up to several hundreds of megaparsecs)it would be undetectable. However its
existence could show up at larger distances.

In the paper we considered only one fluid component filling up the universe.
It is of course oversimplification and is hardly acceptable especially in the
context of negativness of pressure at larger distances. It can be shown, however,
that adding usual dust component into the model do not change relations (16, 17)
significantly (only a constant in (16) will be different) and the qualitative
result will be the same.

It seems that these two different fluid components should effect in the form of
the power spectrum of the CMB anisotropies. Calculation of this power spectrum in
the proposed model will be our next task.
\vspace{8mm}

We thank M.P. D\c abrowski for valuable discussions.

\end{document}